\documentclass[onecolumn,showpacs,floatfix,prb]{revtex4-1}
\usepackage{float}
\usepackage{dcolumn}
\usepackage{hyperref} 
\hypersetup{bookmarks=true,unicode=true,colorlinks=true, urlcolor=blue,linkcolor=blue,citecolor=blue,filecolor=blue}
\usepackage{bussproofs}
\usepackage{graphicx,graphics}
\usepackage{fancyhdr}
\usepackage{amsmath}
\usepackage{subfigure}
\usepackage{epsfig}
\usepackage{amssymb}
\usepackage{bm}
\usepackage{units}
\usepackage{pifont}
\usepackage{textcomp}
\usepackage{gensymb}
\usepackage{enumerate}
\usepackage{color}
\usepackage{physics}

\begin{document}

\title{On the role of local many-body interactions on the thermoelectric  \\ 
properties of fullerene junctions}

\author{C.A. Perroni, V. Cataudella} 
\address{CNR-SPIN and Physics Department "Ettore Pancini", 
Universita' degli Studi di Napoli Federico II, \\ 
Complesso Universitario Monte S. Angelo, Via Cintia, I-80126 Napoli, Italy}

\begin{abstract}
The role of local electron-vibration and electron-electron interactions on the thermoelectric properties of molecular junctions is theoretically analyzed focusing on devices based on
fullerene molecules.  A self-consistent adiabatic approach is used in order to obtain a non-perturbative treatment of the electron coupling to low frequency vibrational modes, such as those of the 
molecule center of mass between metallic leads. The approach incorporates also the effects of strong electron-electron interactions between molecular degrees of freedom within the Coulomb blockade regime. The analysis is based on a one-level model which takes into account the relevant transport level of fullerene and its alignment to the chemical potential of the leads. 
We demonstrate that  only the combined effect of local electron-vibration and electron-electron interactions is able to predict the correct behavior of both the charge conductance and the Seebeck coefficient in very good agreement with available experimental data.
\end{abstract}

\maketitle

\section{Introduction}
In recent years, the field of molecular thermoelectrics has attracted a lot of attention \cite{Cuevas2010,16,17,Yee2011,19,20,21,22,23,PerroniReview,Cui,Park}.
One of the aims is to improve the thermoelectric efficiency of nanoscale devices by controlling the electronic and vibrational degrees of freedom of the molecules. 
Moreover, useful information on charge and energy transport mechanisms can be extracted by studying the thermoelectric
properties of molecular junctions \cite{17,Yee2011,24,Cuevas2010,26}. In addition to the charge conductance $G$, the Seebeck coefficient $S$ is typically measured in these devices.  
Measurements in junctions with fullerene ($C_{60}$) have found a high value of thermopower (of the order or even smaller than $-30 \mu V/K$) \cite{Yee2011}.
Understanding the thermopower  is also important for helping advances in thermoelectric performance of large-area molecular junctions \cite{Park1,Kang}.
Moreover, recently, the application of an Al gate voltage at $Au-C_{60}-Au$ junction has allowed to achieve the electrostatic control of charge conductance and thermopower with unprecedented control \cite{Reddy2014}.  
However, the precise transport mechanisms affecting both $G$ and $S$ remain elusive in these kinds of measurements. Finally, due to experimental challenges \cite{16,27,Wang2007,29}, only very the thermal conductance of 
single-molecule junctions has been fully characterized \cite{Reddy2019}.

In molecular junctions, relevant contributions to the thermoelectric properties typically result from intramolecular electron-electron and electron-vibration interactions \cite{Cuevas2010,32}. 
An additional source of coupling between electronic and vibrational degrees of freedom is also provided by the center of mass oscillation of the molecule between the metallic leads \cite{Park2000}. 
Different theoretical techniques  \cite{Cuevas2010, 32} have been used to study the effects of local many-body interactions which affect the thermoelectric transport  
properties \cite{21,22,23,36,37,Liu2010,39,41} in a significant way. 

In devices with large molecules such as fullerenes or carbon nanotube quantum dots, a non-perturbative treatment of electron-vibration coupling can be obtained within an adiabatic
approach which is based on the slowness of the relevant vibrational modes in comparison with the fast electron dynamics 
\cite{44,45,46,Bode2012,Nocera2011,Nocera2012,Nocera2013,Perroni2014,Perroni2013,Perroni2014b,NoceraReview}. The adiabatic approach can also include a strong
Coulomb repulsion allowing the self-consistent calculation of thermoelectric  properties of massive molecules, such as fullerenes, within the Coulomb blockade regime \cite{Perroni2014a}.

In this paper, the thermoelectric properties of a molecular junction are analyzed focusing on  the role of electron-electron and electron-vibration interactions.  An adiabatic approach 
developed in the literature  takes into account the interplay between the low frequency center of mass oscillation of the molecule and the electronic
degrees of freedom within the Coulomb blockade regime \cite{Perroni2014a}. Parameters appropriate for junctions with $C_{60}$ molecules are considered in this paper. In particular, a one-level model is taken into account since it describes the relevant transport level of fullerene and its alignment to the chemical potential of the metallic leads.  

Aim of this paper is to thoroughly investigate both the charge conductance and the Seebeck coefficient since accurate experimental data are available for $Au-C_{60}-Au$ junction in \cite{Reddy2014} as a function of the voltage gate. We show that an accurate description of the transport properties is obtained in the intermediate regime for the electron-vibration coupling 
and in the strong coupling regime for the electron-electron interaction. Moreover, we point out that only the combined effect of electron-vibration and electron-electron interactions is able to predict the correct behavior of both the charge conductance and the Seebeck coefficient finding a very good agreement with available experimental data. 

The paper is divided as follows. In Section \ref{Model} a very general model for many electronic levels and multiple vibrational degrees is considered and the adiabatic approach is exposed;   
in Section \ref{One-level} the one-level model is presented; in Section \ref{Results} the theoretical results are presented together with the precise comparison with experimental data; finally, in 
Section \ref{Conclusions} conclusions and final discussions are given.

\section{Model and method} \label{Model}
 
In this section, we introduce a general Hamiltonian for a multilevel molecule including many-body interactions between molecular degrees of freedom: the local electron-electron interaction and the local electron coupling to  molecular vibrational modes.  The model simulates also the coupling of the molecule to two leads in the presence of a  finite bias voltage and temperature gradient. The total Hamiltonian of the system is
\begin{equation}\label{Htot}
\hat{\cal H}= \hat H_{mol} + \hat{H}_{leads} + {\hat H}_{leads-mol},
\end{equation}
where $\hat{H}_{mol}$ is the Hamiltonian describing the molecular degrees of freedom, $\hat H_{leads}$ the leads' degrees of freedom and  ${\hat H}_{leads-mol}$ the coupling between 
molecule and leads. 

In this paper,  we assume, as usual in the field of molecular junctions, that the electronic and vibrational degrees of freedom in the metallic leads are not interacting \cite{Cuevas2010,Haug2008}, therefore, the electron-electron and electron-vibration interactions are effective only on the molecule. In Equation (\ref{Htot}), the molecule Hamiltonian $\hat{H}_{mol}$ is
\begin{equation}\label{Hdot}
\hat{H}_{mol}= \sum_{m,l,\sigma}{\hat
c^{\dag}_{m,\sigma}}\varepsilon^{m,l}_{\sigma}{\hat
c_{l,\sigma}}+U \sum_{m,l}{\hat n^{\dag}_{m,\uparrow}}{\hat
n_{l,\downarrow}}+ \hat{H}_{osc} + \hat{H}_{int},
\end{equation}
where $c_{m,\sigma}$ ($c_{m,\sigma}^{\dag}$) is the standard
electron annihilation (creation) operator for electrons on the molecule
levels with spin $\sigma=\uparrow,\downarrow$, where indices $m,l$ can assume positive integer values with a maximum $M$ indicating the total number of electronic levels in the molecule. The matrix $\varepsilon^{m,l}_{\sigma}$ is assumed diagonal in spin space, $\hat{n}_{l,\sigma}=c^{\dag}_{l,\sigma}c_{l,\sigma}$ is the
electronic occupation operator relative to level $l$ and spin $\sigma$, $U$ represents the Coulomb Hubbard repulsion between electrons. We assume that only the diagonal part of the matrix
$\varepsilon^{m,l}_{\sigma}$ is nonzero and independent of the spin:
$\varepsilon^{m,m}_{\sigma}=\varepsilon_{m}$,  where 
$\varepsilon_{m}$ are the energies of the molecule levels. 

In Equation (\ref{Hdot}), the molecular vibrational degrees of freedom are described by
the Hamiltonian
\begin{equation}\label{Hosc}
\hat{H}_{osc}= \sum_{s}{\hat{p}^{2}_{s}\over
2m_{s}}+V({\boldsymbol X}),
\end{equation}
where $s=(1,..,N)$, with $N$ being the total number of vibrational modes, $m_{s}$ is the effective mass associated with the
$s$-th vibrational mode, and $\hat{p}_{s}$ is its momentum operator. Moreover,
$V({\boldsymbol X})={1\over2}\sum_{s}k_{s}{\hat x}_{s}^{2}$ is the
harmonic potential (with $k_{s}$ the spring constants, and
the oscillator frequencies $\omega^{s}_{0}=\sqrt{k_{s}/m_{s}}$), 
$\hat{x}_{s}$ is the displacement operator of the vibrational mode $s$, and ${\boldsymbol X}=(\hat{x}_{1},..,\hat{x}_{N})$ indicates all the displacement operators. 

In Equation (\ref{Hdot}), the electron-vibration coupling $\hat{H}_{int}$ is assumed linear in the vibrational displacements and proportional to the electron level occupations
\begin{equation}\label{Hint}
\hat{H}_{int}= \sum_{s,l}\lambda_{s,l}\hat{x}_{s}\hat{n}_{l},
\end{equation}
where $s=(1,..,N)$ indicates the vibrational modes of the molecule, $l=(1,..,M)$ denotes its electronic levels, $\hat{n}_{l}=\sum_{\sigma} n_{l,\sigma} $ is the
electronic occupation operator of the level $l$,  and $\lambda_{s,l}$ is a matrix representing the electron-vibrational coupling.

In Equation (\ref{Htot}), the Hamiltonian of the electron leads is given by
\begin{equation}\label{Hleads}
\hat{H}_{leads}=
\sum_{k,\alpha,\sigma} \varepsilon_{k,\alpha}
{\hat c^{\dag}_{k,\alpha,\sigma}}{\hat c_{k, \alpha,\sigma}},
\end{equation}
where the operators ${\hat c^{\dag}_{k,\alpha,\sigma}} ({\hat
c}_{k, \alpha,\sigma})$ create (annihilate) electrons with
momentum $k$, spin $\sigma$, and energy
$\varepsilon_{k,{\alpha}}=E_{k,\alpha}-\mu_{\alpha}$ in the left
($\alpha=L$) or right ($\alpha=R$) leads. The
left and right electron leads will be considered as thermostats in
equilibrium at the temperatures $T_L=T+\Delta T/2$ and $T_R=T-\Delta T/2$, respectively, with $T$ the average temperature and $\Delta T$ temperature difference.
Therefore, the left and right electron leads are characterized by the free
Fermi distribution functions $f_{L}(E)$ and $f_{R}(E)$,
respectively, with $E$ the energy. The difference of the electronic chemical potentials in the leads provides the bias
voltage $V_{bias}$ applied to the junction: $\mu_{L}=\mu+e
V_{bias}/2$, $\mu_{R}=\mu-e V_{bias}/2$, with $\mu$ the average
chemical potential and  $e$ the electron charge. In this paper, we will focus on the regime of linear response, that involves very small values of bias voltage $V_{bias}$ and temperature  
$\Delta T$.

Finally, in Equation (\ref{Htot}), the coupling between the molecule and the leads is described by
\begin{equation}\label{Hmol-leads}
\hat{H}_{mol-leads}=
\sum_{k,\alpha,m,\sigma}(V^{m}_{k,\alpha}{\hat
c^{\dag}_{k_{\alpha},\sigma}}{\hat c_{m,\sigma}}+ h.c.),
\end{equation}
where the tunneling amplitude between the molecule and a
state $k$ in the lead $\alpha$ has the amplitude $V^{m}_{k,\alpha}$. 
For the sake of simplicity, we will suppose that the density of
states $\rho_{k,\alpha}$ for the leads is flat within the
wide-band approximation: $ \rho_{k,\alpha} \mapsto \rho_{\alpha}$,
$V_{k,\alpha}^{m} \mapsto V_{\alpha}^{m}$. Therefore, the full hybridization width matrix of the molecular
orbitals is $\Gamma^{m,n}=\sum_{\alpha } 
\Gamma^{m,n}_{\alpha} =\sum_{\alpha } 
\Gamma^{m,n}_{\alpha}$, with the tunneling rate
$\Gamma^{m,n}_{\alpha}=2\pi\rho_{\alpha}V^{m*}_{\alpha}V^{n}_{\alpha}$.
In this paper, we will consider the symmetric configuration:
$\boldsymbol{\Gamma}_{L} = \boldsymbol{\Gamma}_{R} = \boldsymbol{\Gamma}/2$,
where, in the following, bold letters indicate matrices.

In this paper, we consider the electronic system coupled to slow vibrational modes:
$\omega^{s}_{0} \ll \Gamma^{m,n}$, for each $s$ 
and all pairs of $(m,n)$. In this limit, we can treat the mechanical
degrees of freedom as classical, acting as  slow classical fields
on the fast electronic dynamics.  Therefore, the electronic
dynamics is equivalent to a multi-level problem
with energy matrix
$\varepsilon^{m} \rightarrow \varepsilon^{m}+\lambda_{m} x_{m}$,
where $x_{m}$ are now classical displacements  \cite{Bode2012,NoceraReview}. This is called in the literature adiabatic approximation for vibrational
degrees of freedom. 

Within the adiabatic approximation, one gets Langevin self-consistent equations for the vibrational modes of the molecule \cite{Nocera2011,NoceraReview}
\begin{equation}\label{Langevin}
m_{s}\ddot{x}_{s}+k_{s}x_{s}=F_{s}(t)\mathcal+\xi_{s}(t),
\end{equation}
where the generalized force $F_{s}$ is due to the effect of all
electronic degrees of freedom through the electron-vibration coupling \cite{Bode2012,NoceraReview}:
\begin{equation}
F^{el}_s(t)=Tr[i\lambda_{s}{\boldsymbol G}^{<}(t,t)],
\end{equation}
with the trace \`{}\`{}Tr\'{}\'{}, taken over the molecule levels,
defined in terms of the lesser molecular matrix Green's function ${\boldsymbol
G}^{<}(t,t')$  with
matrix elements $ G^{<}_{m,l}(t,t')=i\langle
c^{\dag}_{m,\sigma}(t)c_{l,\sigma}(t')\rangle$. Quantum electronic density fluctuations  on the oscillator motion are responsible for the fluctuating force $\xi_{s}(t)$ in Eq. (\ref{Langevin}) which will be derived below together with generalized force. 

In deriving equations within the adiabatic approximation \cite{NoceraReview}, next, for the sake of simplicity, we do not include explicitly the effect of the Coulomb
repulsion on the molecule Hamiltonian. In the next section, we will show that, in the case of a single level molecule with large repulsion $U$, 
the adiabatic approach works exactly as in the non-interacting case provided that each Green's function pole is treated as a non interacting level \cite{Perroni2014a}.

In our notation  ${\boldsymbol G}$ denotes full Green's functions, while ${\mathcal G}$ denotes the strictly adiabatic
(or frozen) Green's functions which are calculated at a fixed value of ${\boldsymbol X}$. Starting from the Dyson equation \cite{Haug2008,Bode2012,NoceraReview},
the adiabatic expansion for the retarded Green's function  ${\boldsymbol G}^{R}$ is given by
\begin{equation}\label{adiabaticGR}
{\boldsymbol G}^{R}\simeq {\mathcal
G}^{R}+{i\over2}\big(\partial_{E}{\mathcal G}^{R}(\sum_{s}
\lambda_{s} \dot{x}_{s} ){\mathcal G}^{R}-{\mathcal G}^{R}(\sum_s
\lambda_{s} \dot{x}_{s})\partial_{E}{\mathcal G}^{R}\big),
\end{equation}
where ${\mathcal G}^{R}(E,{\boldsymbol X})$ is the strictly adiabatic (frozen) retarded Green's function including the coupling with the leads
\begin{equation}
{\mathcal G}^{R}(E,{\boldsymbol X})=[E-{\boldsymbol \varepsilon}({\boldsymbol
X})-{\boldsymbol \Sigma}^{R,leads}]^{-1}, 
\end{equation}
${\boldsymbol \varepsilon}({\boldsymbol X})$ represents the matrix $\varepsilon^{m,l}_{\sigma}+\sum_{s}\lambda_{s}x_{s}\delta_{l,m}$ and ${\boldsymbol \Sigma}^{R,leads}=\sum_{\alpha}{\boldsymbol \Sigma}^{R,leads}_{\alpha}$ is the total self-energy due to the coupling between the molecule and the leads. For the lesser Green's function ${\boldsymbol G}^{<}$, the adiabatic approximation involves 
\begin{eqnarray}\label{Glesadia}
{\boldsymbol G}^{<}&\simeq&{\mathcal
G}^{<}+{i\over2}\Big[\partial_{E}{\mathcal G}^{<}\big(\sum_{s}
\lambda_{s}\dot{x}_{s} \big){\mathcal
G}^{A}-{\mathcal G}^{R}\big(\sum_s \lambda_{s}
\dot{x}_{s} \big)\partial_{E}{\mathcal
G}^{<}\nonumber\\&&+\partial_{E}{\mathcal G}^{R}\big(\sum_{s}
\lambda_{s}\dot{x}_{s} \big){\mathcal
G}^{<}-{\mathcal G}^{<}(\sum_{s} \lambda_{s}
\dot{x}_{s})\partial_{E}
{\mathcal G}^{A}\Big],
\end{eqnarray}
with ${\mathcal G}^{<} = {\mathcal
G}^{R}{\boldsymbol\Sigma}^{<}{\mathcal G}^{A}$.

The electron-vibration induced forces at the zero order of the
adiabatic limit (${\boldsymbol G}^{<}\simeq {\mathcal G}^{<}$) 
are given by
\begin{equation}\label{force0}
F_{s}^{el(0)}({\boldsymbol X})=-k_{s}x_{s}-\int {dE\over 2\pi i}
tr[\lambda_{s}{\mathcal G}^{<}].
\end{equation}
The leading order correction to the lesser Green's function ${\boldsymbol G}^{<}$ provides a term
proportional to the vibrational velocity 
\begin{equation}\label{force1}
F_{s}^{el(1)}({\boldsymbol X})=-\sum_{s'}\theta_{s,s'}({\boldsymbol
X}) \dot{x}_{s'},
\end{equation}
where the tensor ${\boldsymbol \theta}$ can be split into symmetric and
anti-symmetric contributions \citep{Bode2012}: ${\boldsymbol \theta}={\boldsymbol \theta}^{sym} +
{\boldsymbol \theta}^{a}$, where we have introduced the notation
$\{C_{s,s'}\}_{sym,a}={1\over2}\{C_{s,s'}\pm C_{s',s}\}_{sym,a}$ for
symmetric and anti-symmetric parts of an arbitrary matrix ${\boldsymbol
C}$. Indeed, there is a dissipative term ${\boldsymbol
\theta}^{sym}$ and an orbital, effective magnetic field ${\boldsymbol
\theta}^{a}$ in the space of the vibrational modes.

We can now discuss the stochastic forces $\xi_{s}(t)$ in
Eq. (\ref{Langevin}) within the adiabatic approximation. 
In the absence of electron-electron interactions, the Wick theorem allows to write the noise correlator as
\begin{equation}
\langle \xi^{el}_{s}(t)\xi^{el}_{s'}(t')\rangle=
tr\{\lambda_{s}{\boldsymbol G}^{>}(t,t')\lambda_{s'} {\boldsymbol G}^{<}(t',t)\},
\end{equation}
where ${\boldsymbol G}^{>}(t,t')$ is the greater Green's function with
matrix elements $ G^{>}_{m,l}(t,t')=-i\langle
c_{m,\sigma}(t)c^{\dag}_{l,\sigma}(t')\rangle$.
In the adiabatic approximation, one first substitutes
the full Green's function ${\boldsymbol G}$ by the adiabatic zero-order
Green's function ${\mathcal G}$ and then observes that the
electronic fluctuations act on short time scales only. Therefore, the total forces $\xi_{s}(t)$ are locally correlated in time:
\begin{equation}\label{noise}
\langle \xi^{el}_{s}(t)\xi^{el}_{s'}(t')\rangle\simeq
tr\{\lambda_{s}{\mathcal G}^{>}({\boldsymbol
X},t)\lambda_{s'}{\mathcal G}^{<}({\boldsymbol X},t)\}={\boldsymbol D}({\boldsymbol
X})\delta(t-t'),
\end{equation}
where
\begin{equation}
D_{s,s'}({\boldsymbol X})=\int {dE\over 2\pi}
tr\big\{\lambda_{s}{\mathcal G}^{<}\lambda_{s'}{\mathcal
G}^{>}\big\}_{sym}.
\end{equation}

Once the forces and the noise terms  are calculated, Eq. (\ref{Langevin}) represents a set of non-linear Langevin equations in the unknown $x_{s}$. 
Even for the simple case where only one vibrational degree of freedom is present, 
the stochastic differential equation should be solved numerically in the general non-equilibrium case. 
\cite{Nocera2011,Perroni2013,Perroni2014b} Actually, one can calculate the oscillator distribution
functions $P({\boldsymbol X},{\boldsymbol V})$ (where ${\boldsymbol V}=\dot{\boldsymbol X}=(v_{1},..,v_{N})$),
and, therefore, all the properties of the vibrational modes. Using this
function, one can determine the average $O$ of an electronic or
vibrational observable $O({\boldsymbol X},{\boldsymbol V})$:
\begin{equation}\label{average}
 O=\int \int d {\boldsymbol X} d {\boldsymbol V} P({\boldsymbol X},{\boldsymbol
V}) O({\boldsymbol X},{\boldsymbol V}).
\end{equation}

The electronic observables, such as charge and heat currents, can be evaluated
exploiting the slowness of the vibrational degrees of freedom. In a previous paper \cite{NoceraReview}, we
have discussed the validity of the adiabatic approximation stressing that it is based on the separation between the slow vibrational and fast electronic timescales. 
Actually, physical quantities calculated within the adiabatic approach are very reliable in a large regime of electronic parameters since this self-consistent approach is not perturbative in the 
electron-vibration coupling. Therefore, the approach is able to overcome the limitations of the perturbative theory typically used in the literature \cite{Hsu,Kruchinin}.

\section{One-level model} \label{One-level}

In the remaining part of the paper, we consider the simple case where the molecule is modeled as a single electronic level ($M=1$ in the previous section) locally
interacting with a single vibrational mode ($N=1$ in the previous section). Therefore the focus will be on a molecular level
which is sufficiently separated in energy from other orbitals. In particular, we will analyze the $C_{60}$ molecule where the LUMO energy differs
from the HOMO energy for energies of the order of $1$ eV \cite{Park2000,Lu2003}.  Even when the degeneracy of the LUMO is removed by the contact with metal leads, 
the splitting gives rise to levels which are separated by an energy of the order of a few tenths of eV \cite{Lu2003}. Furthermore, the energy of the molecular
orbital can be tuned by varying the gate voltage $V_{G}$.

One-level transport model has been adopted to interpret experimental data of $C_{60}$ molecular junctions \cite{Reddy2014} neglecting altogether the effect of electron-electron and electron-vibrations interactions. This model 
is clearly valid for energies close to the resonance, therefore it is particularly useful in the case of the experiments in \cite{Reddy2014}  where the molecular energy is tuned around the Fermi energy of the leads.
Moreover, the one-level model has to be used in the regime of low temperatures, therefore temperatures up to room temperature can be considered for the interpretation of experimental data.
Within this model, the energy-dependent transmission function $T(E)$ is assumed to be well approximated by a Lorentzian function:
\begin{equation} \label{transmission}
T(E) = \frac {4 \Gamma^2}
{ (E -\epsilon)^2 + 4\Gamma^2},
\end{equation}
where the molecular level energy $\epsilon$ is taken as
\begin{equation} \label{molevel}
\epsilon= E_0  -\alpha V_G,
\end{equation}
with $E_0$ the energetic separation of the dominant transport level with respect to the chemical potential $\mu$, and $\alpha$ the effectiveness of gate coupling. The expression of $\epsilon$ takes clearly into account the tuning of the molecular level by the gate voltage.
By using  Equation (\ref{transmission}),  in the limit of low temperature of the Landauer-B\"uttiker approach valid in the coherent regime \cite{Cuevas2010,Haug2008}, the gate voltage-dependent
electrical conductance $G$ becomes
\begin{equation} \label{Gcoher}
G=\frac{\partial I}{\partial V_{bias}} (V_{bias}=0,V_G)=G_0 \ T(E=\mu),
\end{equation}
where $G_0=2 e^2/h$ is the quantum  of conductance, with $h$ Planck constant. Moreover, in the same limit, the Seebeck coefficient $S$ is  
\begin{equation} \label{Scoher}
S= -\frac{\pi ^2}{3} \frac{k_B}{|e|} k_B T \frac {\partial \ln T(E=\mu)}{\partial E}=\frac{\pi ^2}{3} \frac{k_B}{|e|} k_B T  
\frac{2 [\mu-\epsilon]}{[(\mu-\epsilon)^2+4 \Gamma^2]},
\end{equation}
where $k_B$ is the Boltzmann constant. We remark that $k_B/|e| \simeq 86.17$  $\mu V/K$ sets the order of magnitude (and, typically, the maximum value in modulus) of the thermopower 
in molecular junctions.

\begin{figure}[t]
\centering
\includegraphics[scale=0.25,angle=-90]{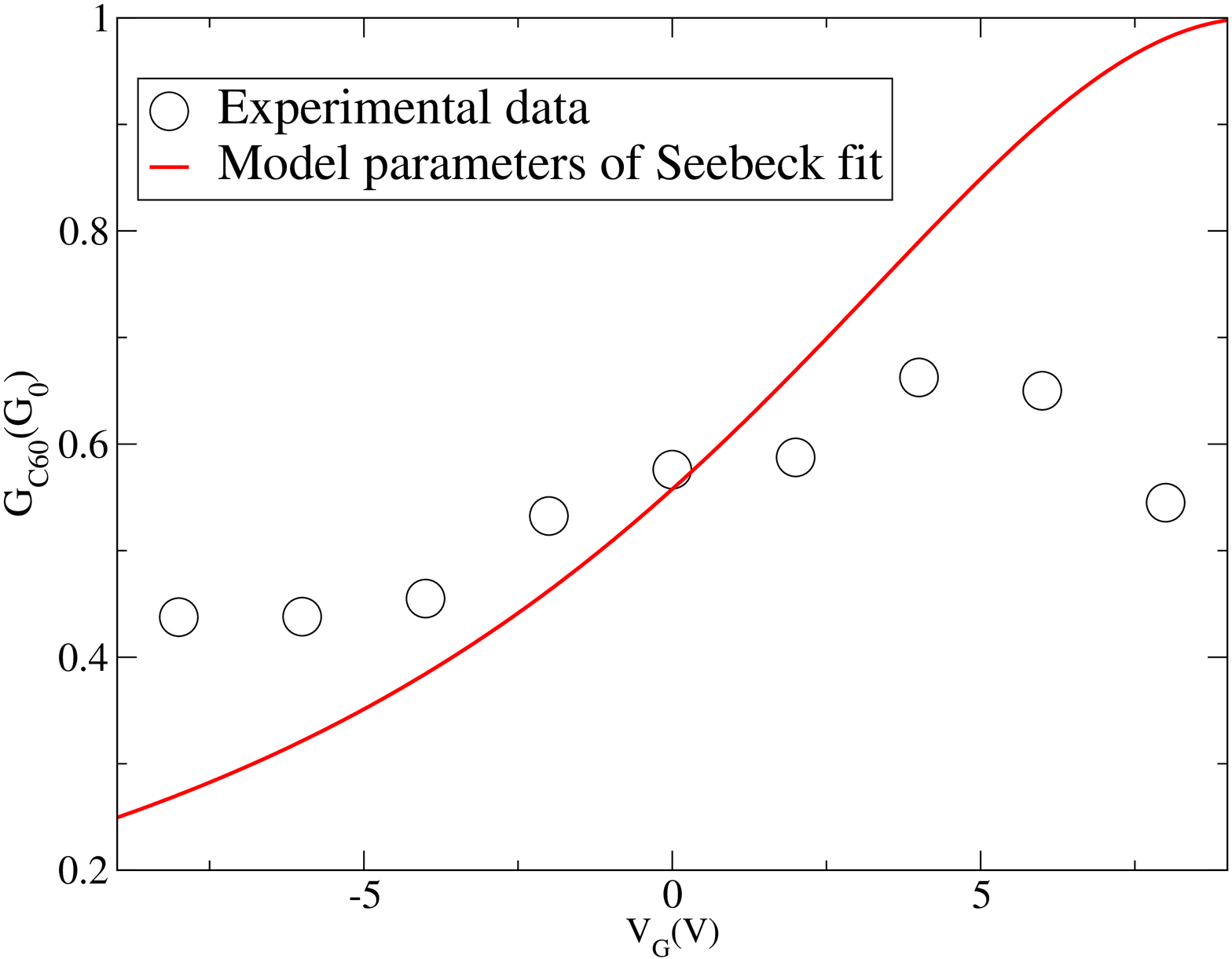}
\includegraphics[scale=0.25,angle=-90]{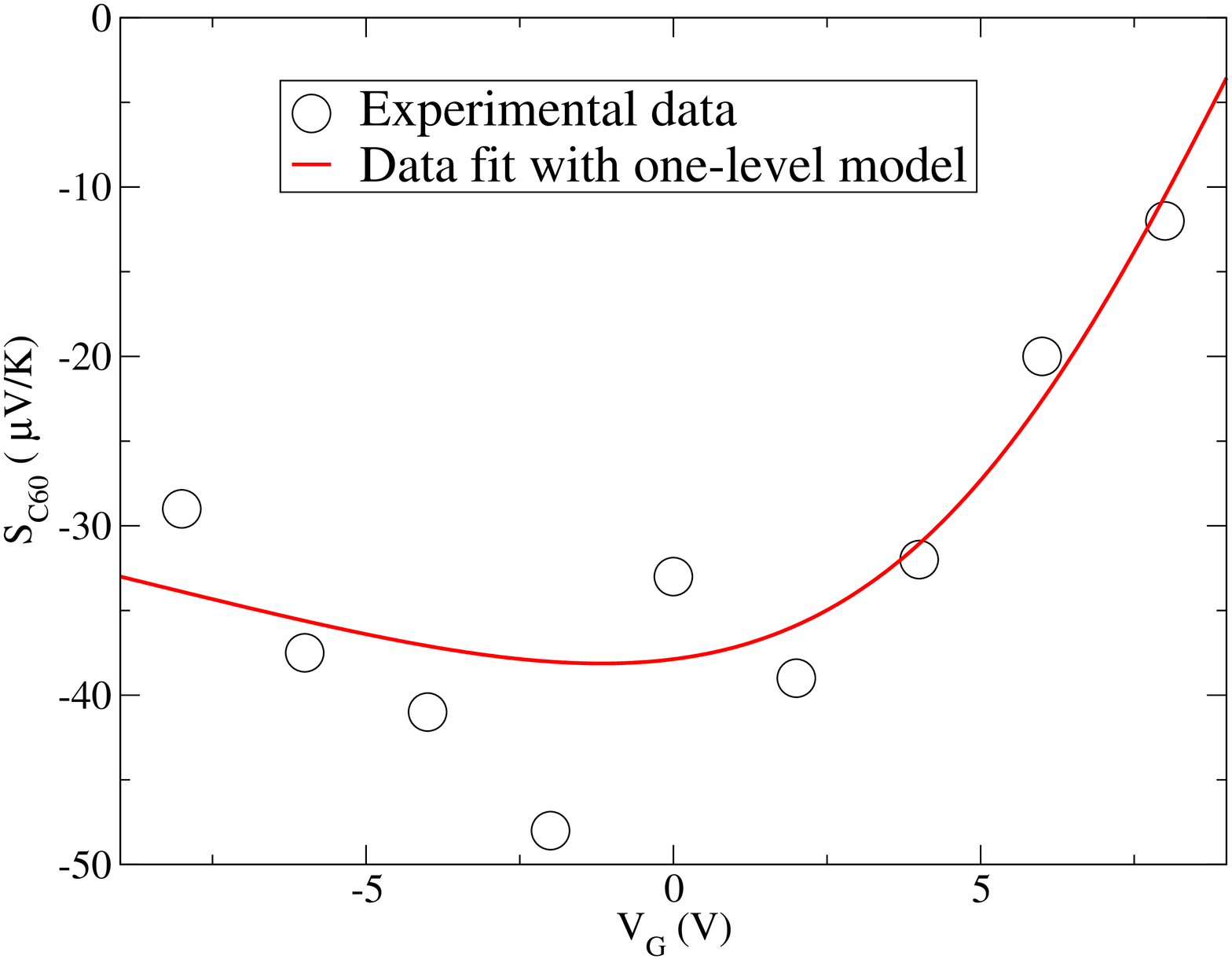}
\caption{Left Panel: Conductance G (in units of  conductance quantum $G_0$) as a function of the gate voltage $V_{G}$ (in units of V) at $T=100$ K from experimental data (black circles, see \cite{Reddy2014} for fullerene $C_{60}$ junctions) and from a curve (red solid line) obtained by using the parameters of the fit to the Seebeck coefficient. Right Panel: Seebeck coefficient S (in units of $\mu K/V$) as a function of the gate voltage $V_{G}$ (in units of V) at $T=100$ K from experimental data (black circles), and from a fit (red solid line). For both, see  \cite{Reddy2014} relative to $C_{60}$ molecular junctions.}
\label{Fig1}
\end{figure}

In the right panel of Figure (\ref{Fig1}), we report the experimental data of Seebeck coefficient S as a function of the gate voltage $V_{G}$ taken from \cite{Reddy2014} for $C_{60}$ junctions. 
The values of  S taken at the temperature $T=100$ K are quite large in modulus for negative gate.  
Moreover, the data  show a marked change as a function of the gate voltage suggesting that the chemical potential is able to cross a level of the molecule. Since the values of $S$ are negative for small values of $V_G$ and are still negative for zero $V_G$, the charge transport is dominated by the lowest unoccupied molecular orbital (LUMO) level of $C_{60}$.  Actually, in order to fit the experimental data shown in the right panel of Figure (\ref{Fig1}), Equation (\ref{Scoher}) has been used getting the positive value 
$E_0-\mu=0.057$ $eV$ \cite{Reddy2014}. For the optimization of the fit, in the same paper  \cite{Reddy2014}, $\Gamma=0.032$ $eV$ and the gate voltage effectiveness $\alpha=0.006$ $eV/V$ are also extracted. These three numerical values put in  Equation (\ref{Scoher}) provide the fit curve shown in the right panel of Figure (\ref{Fig1}). The fit is good, but not excellent. 

In the left panel of Figure (\ref{Fig1}), we report the experimental data of the charge conductance G as a function of the gate voltage $V_{G}$ taken again from experimental data of \cite{Reddy2014} for $C_{60}$ junctions. Even if the temperature is not high ($T=100$ K), the values of G are quite smaller than the conductance quantum $G_0$.  Moreover, if one uses the parameters ($E_0-\mu=0.057$ $eV$, $\Gamma=0.032$ $eV$, and $\alpha=0.006$ $eV/V$) extracted from the Seebeck data in  \cite{Reddy2014} and reproduced in the right panel of  
Figure (\ref{Fig1}), one finds a peak of the conductance for $E_0 -\mu=\alpha V_G$, hence for $V_G \simeq 9$ V. This is in contrast with the peak of $G$ which occurs at $V_G \simeq 5$ V in the experimental data.  If we try to describe the experimental data shown in the left panel of Figure (\ref{Fig1}) by using Equation (\ref{Gcoher}) and the parameters extracted by fitting the Seebeck data, we get the red line reported in the left panel of Figure  (\ref{Fig1}). It is evident that the agreement between theory and data is poor, and, in particular, the maximum observed for $V_G$ around 5 V is not recovered.  We remark that $k_B T \simeq 0.0086$ $eV$ represents the smallest energy scale apart from values of $V_G$ very close to the LUMO level.  Therefore, the quality of the comparison cannot depend on the low temperature expansion used in Equation (\ref{Gcoher}).

In order to improve the interpretation of the experimental data, in this paper, we analyze the role of many-body interactions between molecular degrees of freedom. For example,  experimental measurements have highlighted that the effects of the electron-vibration interactions  are not negligible in junctions with $C_{60}$ molecules and gold electrodes \cite{Park2000,PerroniReview}. In particular, 
experimental results for $C_{60}$ molecules \cite{Park2000}  provide compelling evidence for a sizable coupling between the electrons  and the center of mass vibrational mode.
Indeed, previous studies have shown that a $C_{60}$ molecule is held tightly on gold by van der Waals interactions, which can be expressed by the Lennard-Jones form.  The $C_{60}$-gold binding near the equilibrium position
can be approximated very well by a harmonic potential with angular frequency $\omega_0$. For $C_{60}$ molecules, the center of mass energy $\hbar \omega_0$ has been estimated to be of the order of $5$ meV. 

In this paper,  we will focus on the center of mass mode as the relevant low frequency vibrational mode for the molecule. The center of mass mode is expected to have the lowest angular frequency $\omega_0$ for large molecules. For fullerene, the energy $\hbar \omega_0$ is still smaller than the thermal energy $k_B T$  corresponding to the temperature $T=100$ K fixed for the measurements made in \cite{Reddy2014}.  For $k_B T \ge \hbar \omega_0$, the self-consistent adiabatic approach introduced in the previous section can be used for a non-perturbative treatment of the electron-vibration coupling. Equation (\ref{Langevin}) reduces in this case to a single Langevin equation \cite{Nocera2011,Perroni2014}.
We hereby report the expression for the displacement dependent electronic spectral function $A(E,x)$
\begin{equation}\label{Spec}
A(E,x)= \frac{4 \Gamma}{  (E-\epsilon - \lambda x )^2+4 \Gamma^2}.
\end{equation}
Within these assumptions, in Eq. (\ref{Hdot}), the interaction Hamiltonian ${\hat H}_{int}$ reduces to the same interaction term of the single impurity Anderson-Holstein model \cite{Cuevas2010} and the 
electron-oscillator coupling sets the characteristic polaron energy $E_P$
\begin{equation}
E_{P}={\lambda^2\over2m\omega_{0}^{2}},
\end{equation}
with $m$ mass of the molecule. Actually, an additional electron injected from the leads compresses the $C_{60}$-surface bond shortening the $C_{60}$-surface distance, but not significantly changing the vibrational frequency.
Previous studies \cite{Park2000,PerroniReview} have estimated that the number of vibrational quanta typically excited by the tunnelling electron in fullerene junctions is not large. Therefore, intermediate values of electron-vibration energy $E_P$ corresponding to values comparable with $\Gamma$  are considered relevant for fullerene molecular junctions. Taking the parameters extracted from the experimental data discussed above, $E_P \simeq 0.030$ eV sets the order of magnitude. 
 
In order to improve the analysis of the fullerene molecular junction, in this paper, we study also the role of electron-electron interactions acting onto the molecule. Indeed,  the conductance gap observed in the data of $C_{60}$ molecules can be interpreted  using ideas borrowed from the Coulomb blockade effect \cite{Cuevas2010,Park2000}. Therefore, these features are understood in term of the finite energy required to add (remove) an electron to (from) the molecule. Within the single-level model introduced in the previous section, this charging energy is simulated by fixing the value of the local Hubbard  term U in Equation (\ref{Hdot}). The maximum conductance gap observed in the experimental data \cite{Park2000} indicates that the charging energy of the $C_{60}$ molecule can be around $0.27$ eV, therefore experiments set the order of magnitude $U \simeq 0.3$ eV. 

In order to include the Coulomb blockade effect within the  adiabatic approach  discussed previously, we generalize it to the case in which  the electronic level can be double occupied and a strong Coulomb repulsion $U$ is added together with the electron-vibration interaction. 
The starting point is the observation that, in the absence of electron-oscillator interaction,
and in the limit where the coupling of the dot to the leads is small $\Gamma<<U$ \cite{Haug2008},
the single particle electronic spectral function is characterized by two spectral  peaks separated by an energy interval equal to $U$. In the adiabatic regime, one can independently perturb each spectral peak of the molecule \cite{Perroni2014a}, obtaining at the zero order of the adiabatic approach
\begin{equation}
A(E,x) = [1-\rho(x)]\frac{4 \Gamma}{(E - \epsilon -\lambda x )^2+4 \Gamma^2}+ 
 \rho(x) \frac{4 \Gamma}{(E - \epsilon -\lambda x -U )^2+4 \Gamma^2}, \label{function}
\end{equation}
where $\rho(x)$ is the electronic level density per spin. In our computational scheme, 
$\rho(x)$ has to be self-consistently calculated for a fixed displacement $x$  of the oscillator through the following integral 
$\rho(x)=\int_{-\infty}^{+\infty} \frac{d E}{2 \pi i}
G^<(E,x)$, with the lesser Green function $G^<(E,x)=\frac{i}{2} [f_L(E)+f_R(E)] A(E,x)$. 
The above approximation is valid if the electron-oscillator interaction is not too large, such that
$\Gamma\simeq E_P<<U$ and the two peaks of the spectral function can  be still resolved \cite{Perroni2014a}. We remark that, in comparison with our previous work \cite{Perroni2014a}, parameters appropriate for junctions with $C_{60}$ molecules are considered in this paper focusing on the temperature $T=100$ K fixed for the measurements made in \cite{Reddy2014}, smaller than the room temperature, where the adiabatic approach can be still adopted. Therefore, the approach is valid in the following parameter regime:  $\hbar\omega_0  \leq k_{B} T< \Gamma \ll U $ \cite{Perroni2014a,NoceraReview}.

Within the adiabatic approach, the actual electronic spectral function $A(E)$ results from the
average over the dynamical fluctuations of the oscillator motion, therefore, as a general observable, it is calculated by using Equation (\ref{average}):
\begin{equation}\label{Specnew}
A(E)=\int_{-\infty}^{+\infty} d x P(x)A(E,x),
\end{equation}
where $P(x)$ is the reduced position distribution function of the oscillator. Notice that, in the absence of electron-electron ($U=0$) and electron-vibration ($E_P=0$) interactions, the spectral function is proportional to the transmission $T(E)$ given in Equation (\ref{transmission}) through the hybridization width $\Gamma$: $T(E)=\Gamma A(E)$.

In the linear response regime (bias voltage $V_{bias} \rightarrow 0^+$ and temperature difference $\Delta T \rightarrow 0^+$), all the electronic transport coefficients can be expressed as integrals of $A(E)$. To this aim, we report  the conductance $G$
\begin{equation}
G=G_0  \Gamma \int_{-\infty}^{+\infty} d E  A(E) \left[ -\frac{\partial
f(E)}{\partial E} \right],\label{condus}
\end{equation}
where $A(E)$ is the spectral function defined in Eq. (\ref{Specnew}),  with $f(E)=1/(\exp{[\beta (E-\mu)]}+1)$ the free
Fermi distribution corresponding to the chemical potential $\mu$ and the temperature $T$, and $\beta=1/k_{B} T$. The Seebeck
coefficient is given by $S=-G_S/G$, where the charge conductance $G$ has been defined in Eq. (\ref{condus}), and 
\begin{equation}
G_S=  G_0 \left( \frac{k_B}{e} \right)  \Gamma 
 \int_{-\infty}^{+\infty} d E \frac{(E-\mu)}{k_B T} A(E) \left[ -\frac{\partial
f(E)}{\partial E} \right]. \label{conducts}
\end{equation}
Then, we will calculate the electron thermal conductance $G_K^{el}=G_Q-T G S^2$, with
\begin{equation}
G_Q=  G_0 \left( \frac{k_B}{e} \right)^2  \Gamma T \int_{-\infty}^{+\infty} d E \left[ \frac{E-\mu}{k_B T} \right]^2 A(E) \left[ -\frac{\partial
f(E)}{\partial E} \right]. \label{conductq}
\end{equation}
Therefore,  in the linear response regime, one can easily evaluate the electronic thermoelectric figure of merit $ZT^{el}$ 
\begin{equation}
ZT^{el}=\frac{G S^2 T}{G_K^{el}},
\end{equation}
which characterizes the electronic thermoelectric conversion. We recall that, in this paper, we will  not consider the addition contribution coming from phonon thermal conductance $G_K^{ph}$.

\section{Results} \label{Results}
In this section, we discuss the thermoelectric properties within the single-level model analyzing the role of the electron-electron and electron-vibration interactions between the molecular degrees of freedom.  We will point out that only the combined effect of these interactions is able to provide a good agreement between experimental data and theoretical calculations.  

\begin{figure}[t]
\centering
\includegraphics[scale=0.25,angle=-90]{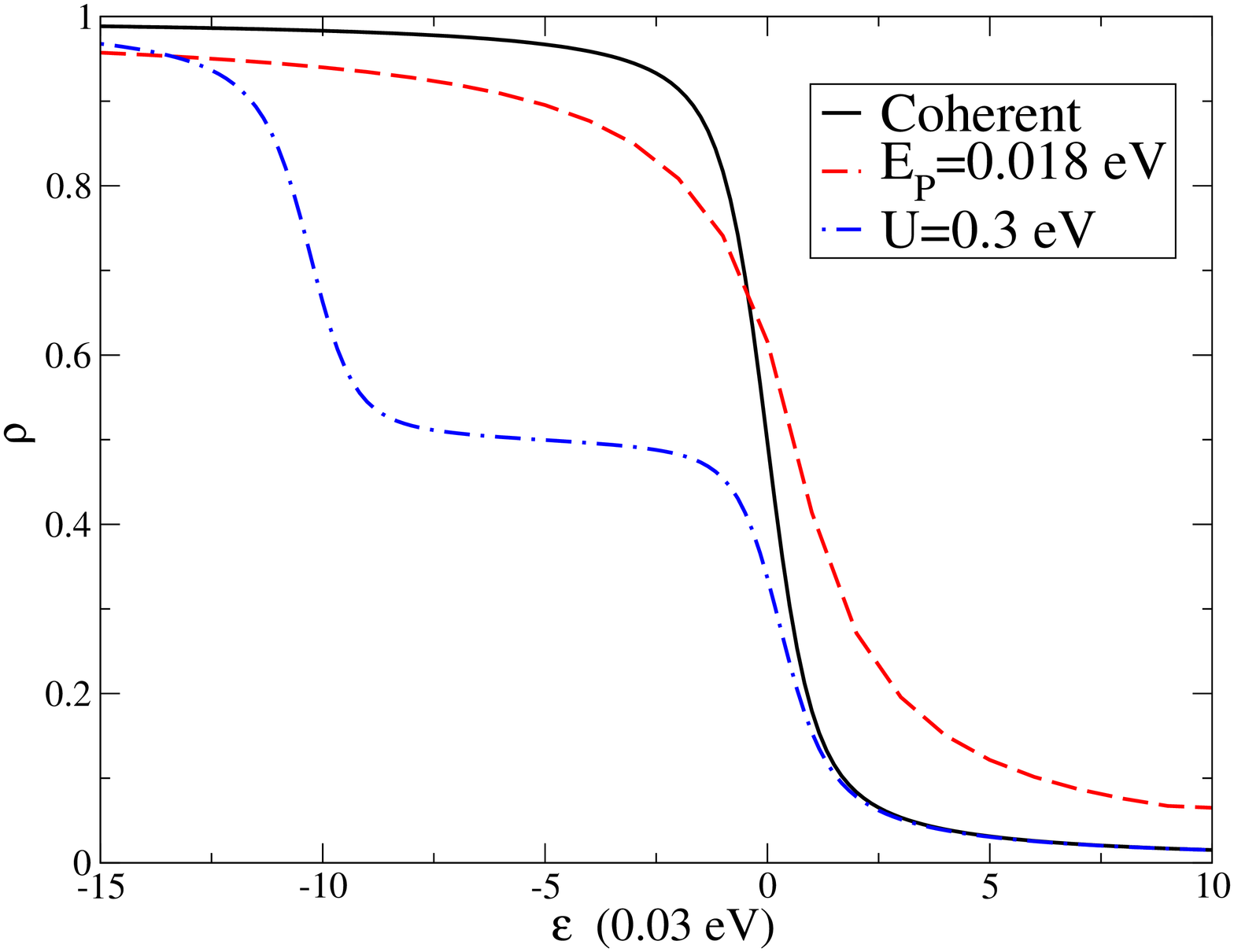}
\includegraphics[scale=0.25,angle=-90]{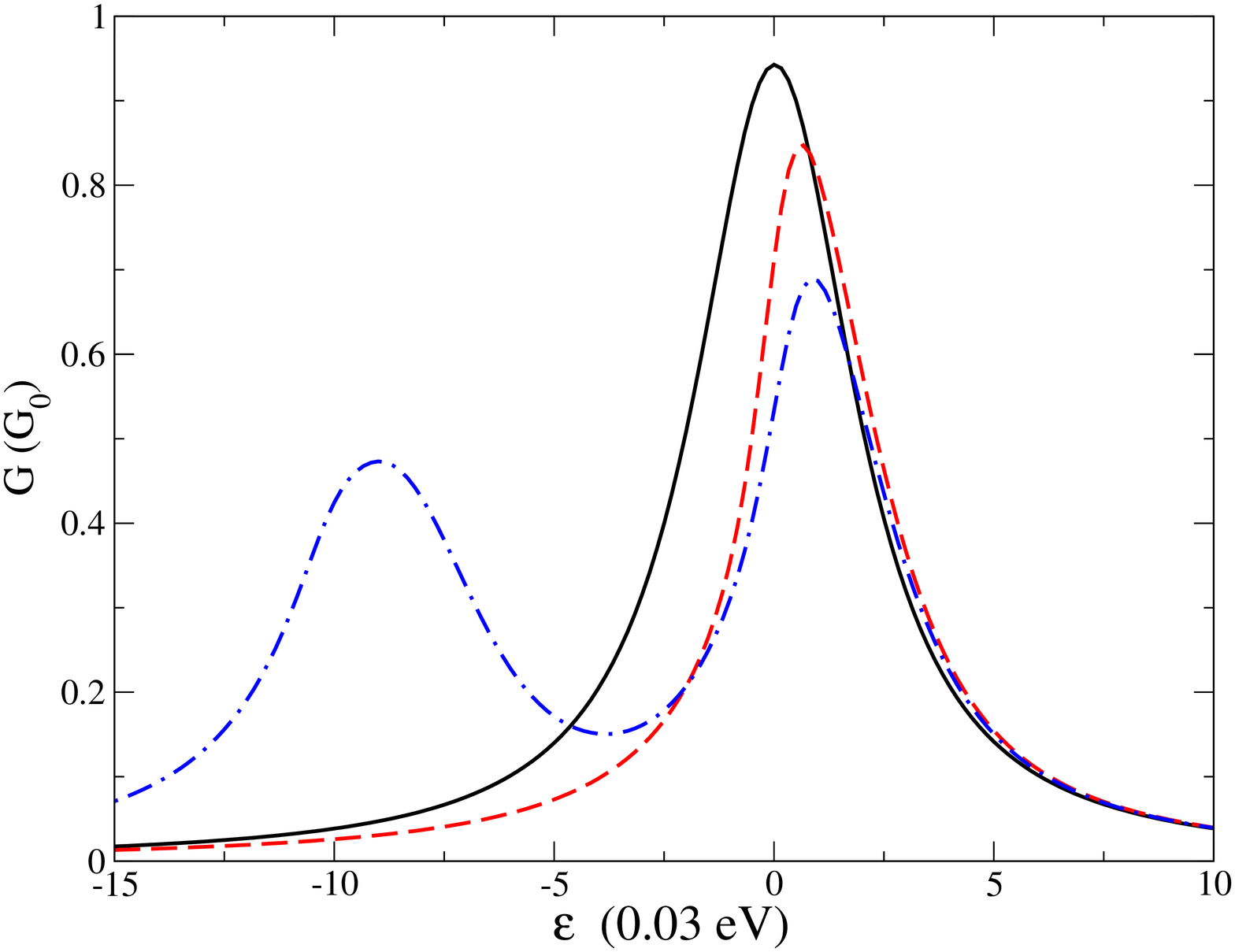} \\
\includegraphics[scale=0.25,angle=-90]{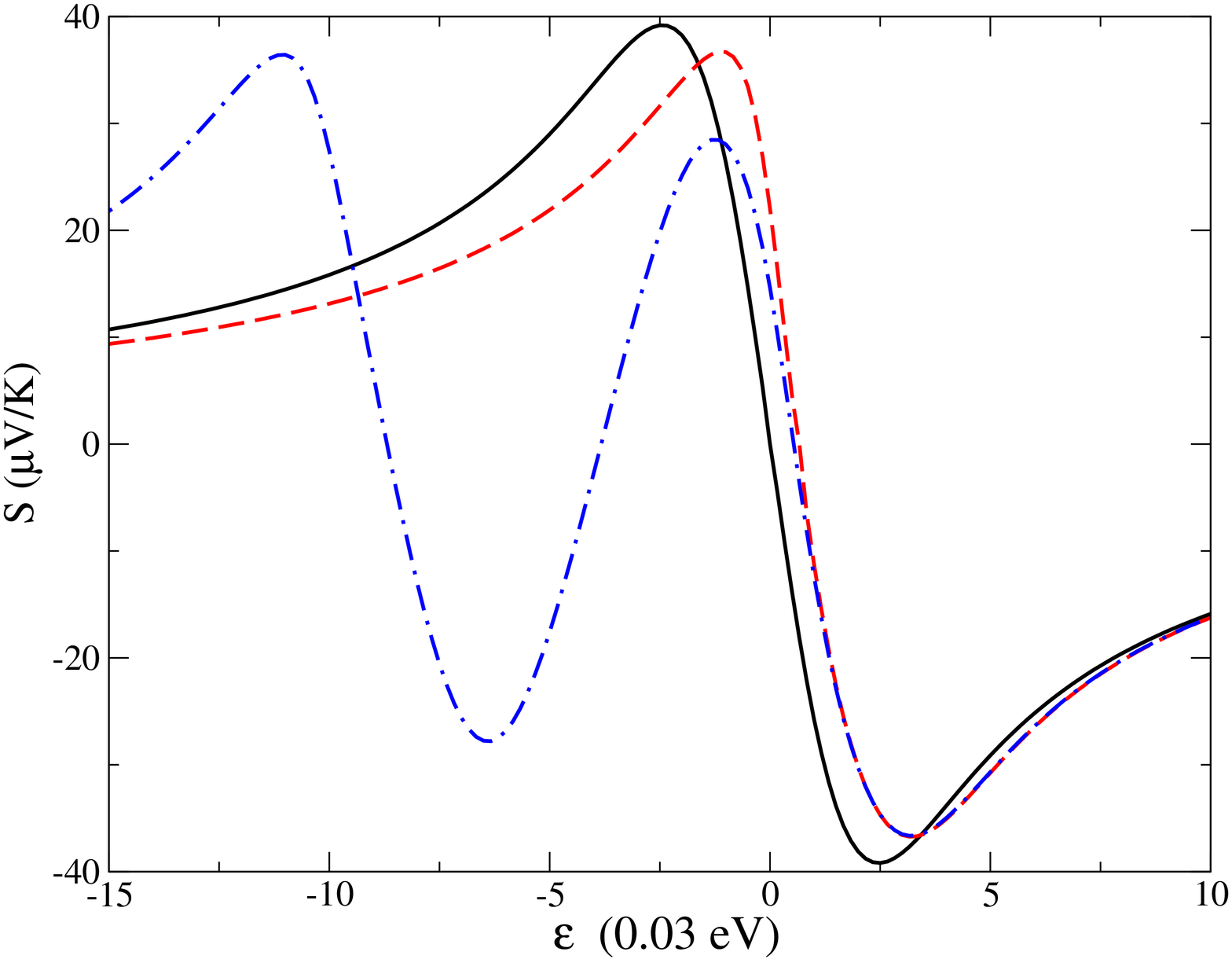}
\includegraphics[scale=0.25,angle=-90]{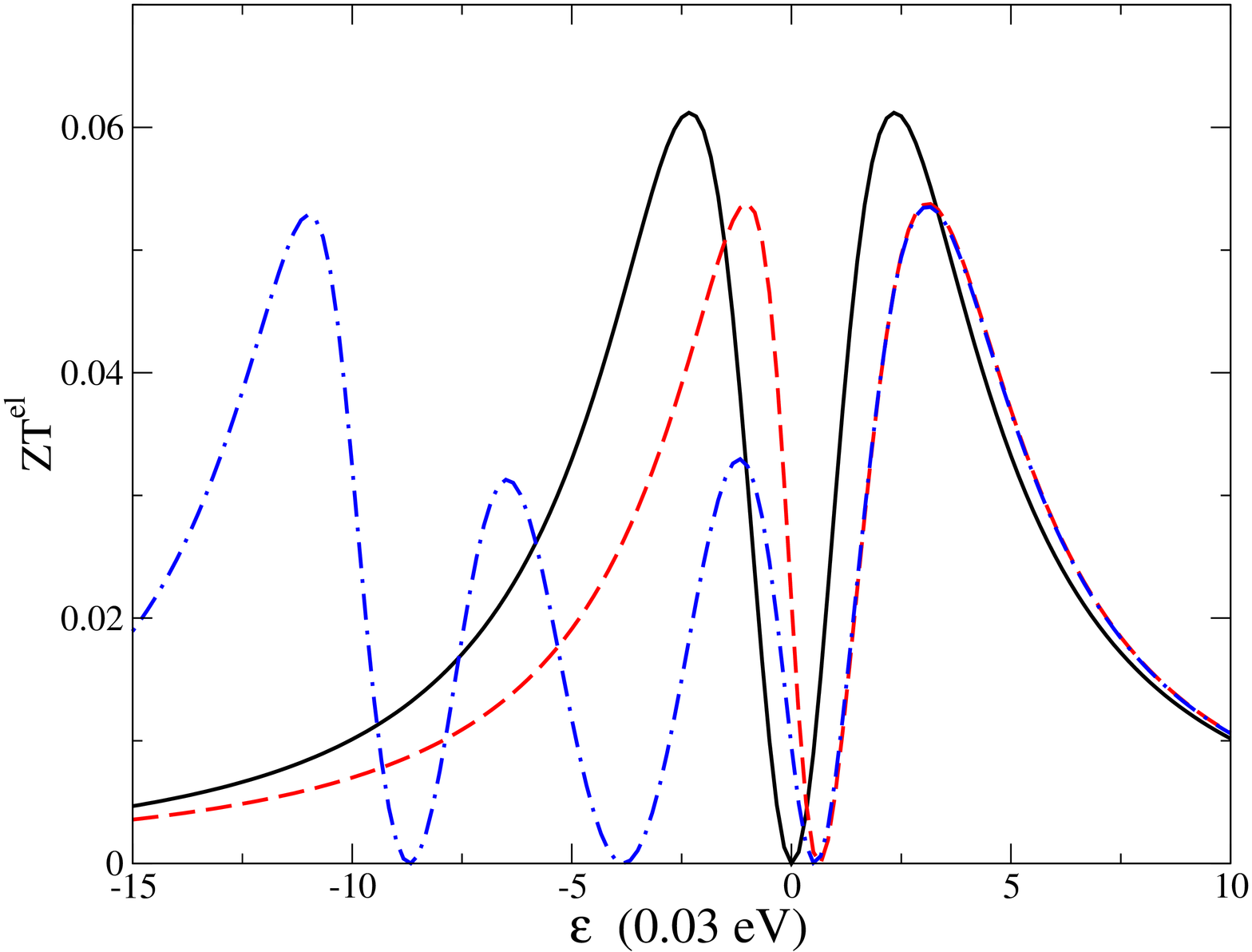}
\caption{Level density $\rho$ in the upper left panel, charge conductance G (in units of the conductance quantum $G_0$) in the upper right panel, Seebeck coefficient $S$ (in units of $\mu V /K$) in the lower left panel, and electronic thermoelectric figure of merit $ZT^{el}$ in the lower right panel as a function of level energy $\epsilon$ (in units of $0.030$ eV) at the temperature $T=100$ K.}
\label{Fig2}
\end{figure}

The level density $\rho$ is shown in the upper left panel of Figure (\ref{Fig2}), the charge conductance G in the upper right panel, the Seebeck coefficient $S$ in the lower left panel, and the electronic thermoelectric figure of merit $ZT^{el}$ in the lower right panel. All the quantities are plotted as a function of level energy $\epsilon$ at the temperature $T=100$ K. For all the quantities, we first analyze the coherent regime (black solid lines in the four panels of  Figure (\ref{Fig2})), that means absence of electron-electron and electron-vibration interactions. Then, we study the effect of a finite electron-vibration coupling $E_P$ (red dash lines in the four panels of  Figure (\ref{Fig2})) focusing on the intermediate coupling regime. Finally, we consider the combined effect of electron-vibration and electron-electron interactions for all the quantities (blue dash-dot lines in the four panels of  Figure (\ref{Fig2})) analyzing the experimentally relevant regime of a large Coulomb repulsion $U$.

The level density $\rho$ per spin reported in the upper left panel of Figure (\ref{Fig2}) shows the expected decreasing behavior with increasing the level energy $\epsilon$. The electron-vibration interaction induces a shift of the curve of about $E_P$.  In the presence  of electron-electron interactions, the behavior is more complex. Actually, in molecular junctions, the strong Coulomb repulsion usually reduces the  electronic charge fluctuations and suppresses the double occupation of the electronic levels \cite{Cuevas2010}. For values of $\epsilon$ smaller than $-U$, the density is closer to unity, while, for $\epsilon$ larger than zero, the density vanishes. For 
$\epsilon$ between $-U$ and $0$, there is a plateau with a value of the density close to $0.5$.  Indeed, these phenomena are characteristic  of Coulomb blockade effects.      

The conductance $G$ is shown in upper right panel of Figure (\ref{Fig2}). At low temperatures, this quantity as a function of the level position $\epsilon$ provides essentially the spectral function of the molecular level. Indeed, in the coherent low temperature regime, $G$ can be directly related to the transmission with a Lorentzian profile. One of the main effects of an adiabatic 
oscillator is to shift the conductance peak towards positive energies proportional to the electron-vibration coupling energy $E_P$. Apparently, another expected effect is the reduction of the peak amplitude.  In fact, electron-vibration couplings on the molecule tend to reduce the charge conduction. As a consequence, electron-vibration couplings induces somewhat longer tails far from the resonance. These features, such as the peak narrowing, are common to other theoretical approaches treating electron-vibration interactions, among which that related to the Franck-Condon blockade \cite{Oppen1}. Actually, in a previous paper \cite{NoceraReview}, we
have successfully compared the results of the adiabatic approximation with those of the Franck-Condon blockade formalism in the low density limit where this latter approach becomes essentially exact \cite{Cuevas2010}.

We note that a finite electron-electron interaction not only suppresses the  electronic conduction for small values of $\epsilon$, but it is also responsible for a second peak centered at $\epsilon \simeq -U$. In fact, there is a transfer of spectral weight from the main peak  to the interaction-induced secondary peak. We stress that these features are compatible with experimental data since conductance gap ascribed to Coulomb blockade effects have been measured in fullerene junctions \cite{Park2000,Cuevas2010}.

We investigate the properties of the Seebeck coefficient $S$ of the junction  in the lower left panel of Figure (\ref{Fig2}). In analogy with the behavior of the conductance, 
the main effect of the electron-vibration interaction is to reduce the amplitude of the Seebeck coefficient. Moreover, the shift  of the zeroes of $S$ is
governed by  the coupling $E_P$ as that of the peaks of $G$. Therefore, with varying the level energy $\epsilon$, if  $G$ reduces its amplitude,
$S$ increases its amplitude in absolute value, and vice versa. This behavior and the values of $S$ are in agreement with experimental data  \cite{Reddy2014,Yee2011}. 
In the Coulomb blockade regime, $S$ shows a peculiar oscillatory behaviour as a function of the energy $\epsilon$, with several positive peaks and negative dips. The energy distance between the peaks (or the dips) is governed by the Hubbard term $U$. Even in this regime, the Seebeck coefficient $S$ is negligible
for the level energies where the electronic conductance presented the main peaks, that is at $\epsilon \simeq 0$ and $\epsilon \simeq -U$. This property turns out to be a result of the 
strong electron-electron interaction $U$ \cite{Liu2010}. In any case, close to the resonance (zero values of the level energy $\epsilon$), the conductance looks more sensitive to many-body interactions, while the Seebeck coefficient appears to be more robust.

The electronic conductance $G$, Seebeck coefficient $S$, and electron thermal conductance $G_{K}^{el}$ combine in 
giving an electronic figure of merit $ZT^{el}$. This latter quantity is shown in lower right panel of Figure (\ref{Fig2}) at the temperature $T=100$ K. We stress that, due to the low value of the temperature, the quantity  $ZT^{el}$ does not show values comparable with unity. However, it is interesting to analyze the effects of many-body interactions on this quantity. A finite value of  the electron-vibration coupling $E_P$ leads to a reduction of the height of the figure of merit peaks. By the way, the position of the peaks in $ZT^{el}$ roughly coincides with the position of the peaks and dips of the Seebeck coefficient S. Finally,  the electron-electron interactions tend to reduce the amplitude and to further shift the peaks of the figure of merit.

\begin{figure}[t]
\centering
\includegraphics[scale=0.27,angle=-90]{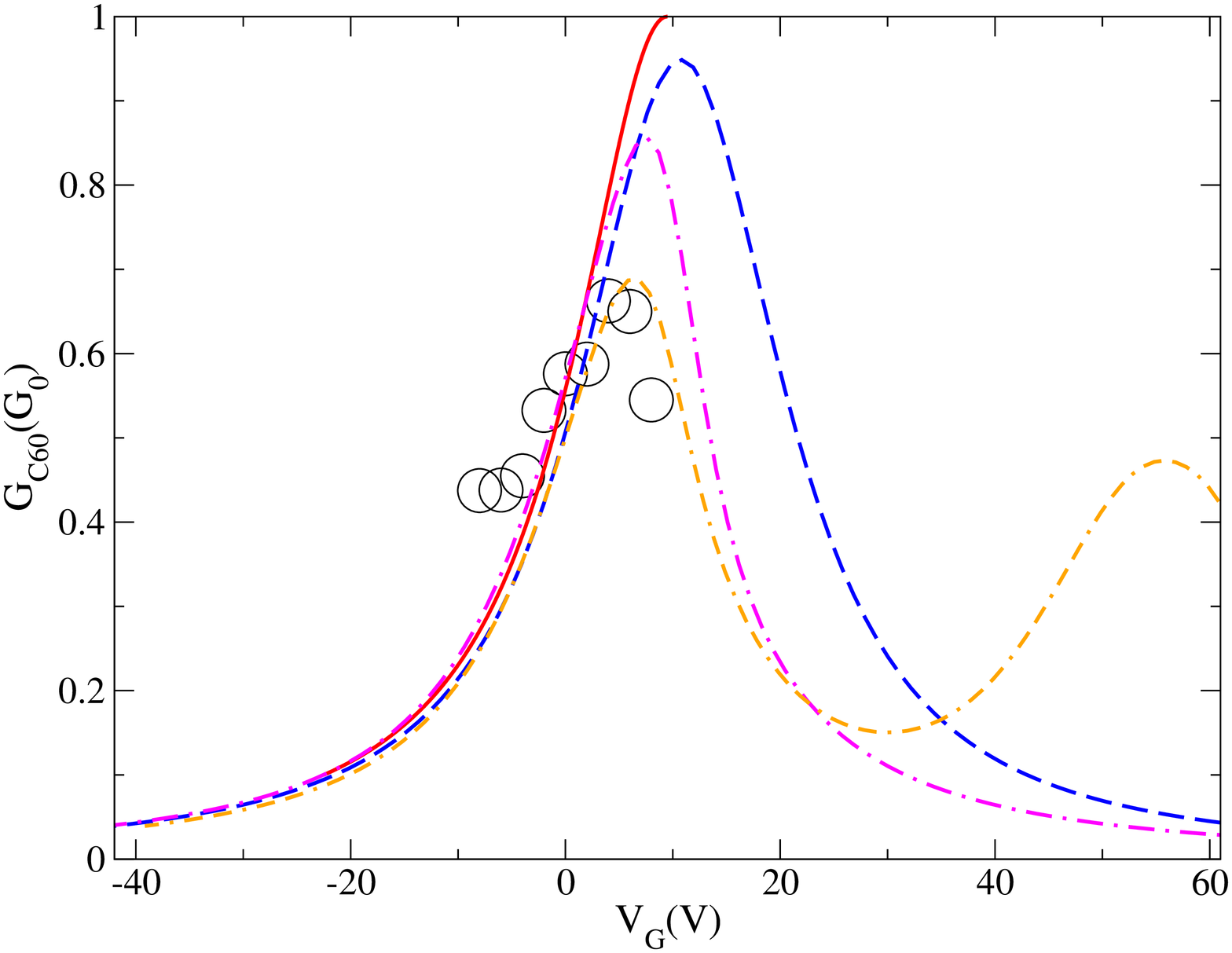}
\includegraphics[scale=0.27,angle=-90]{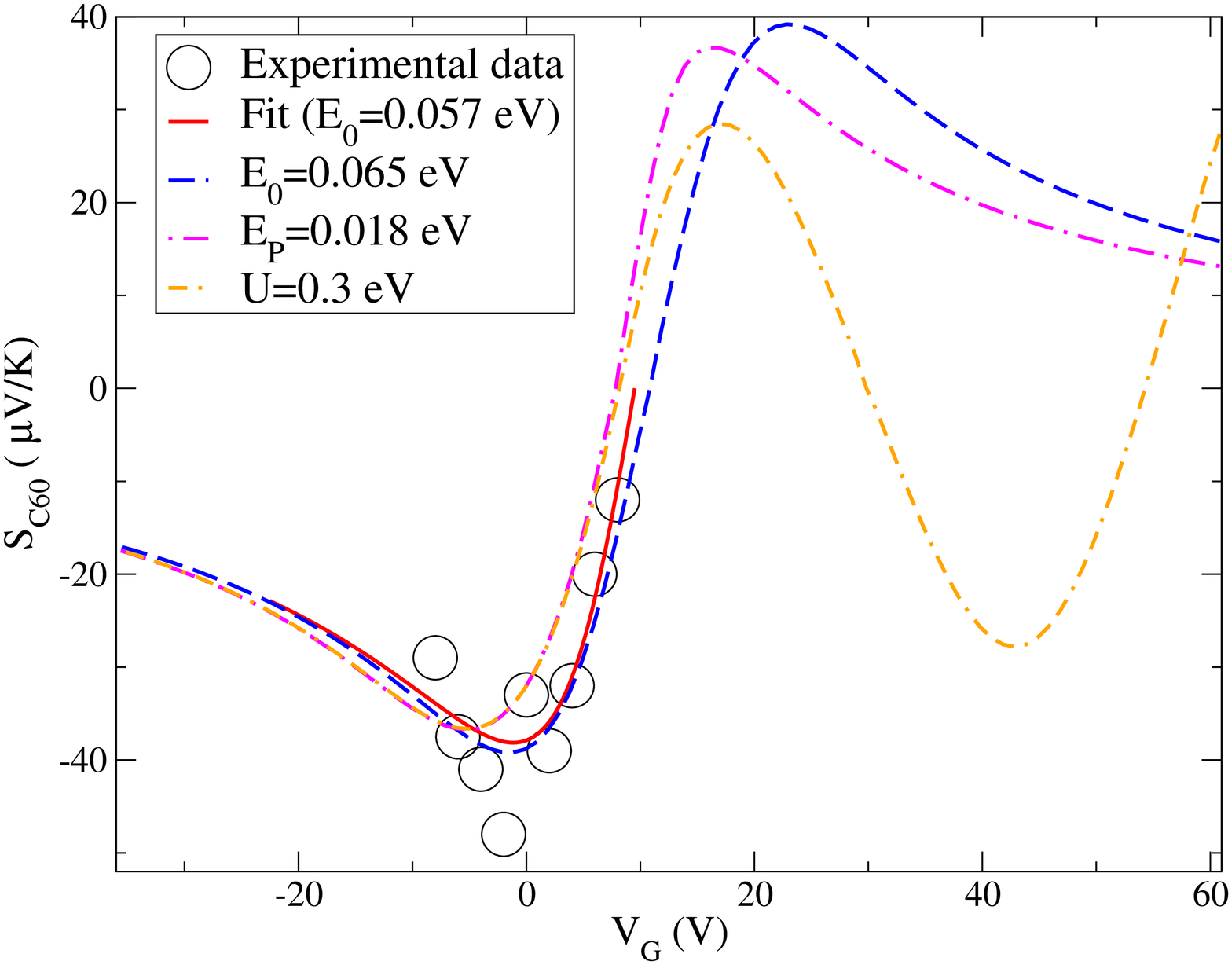}
\caption{Charge conductance G (in units of the conductance quantum $G_0$) in the left panel, Seebeck coefficient $S$ (in units of $\mu V /K$) in the right panel as a function of gate voltage  
$V_G$ (in units V) at the temperature $T=100$ K: experimental data (black circles), data fit (red solid line) corresponding to one-level model with energy $E_0-\mu=0.057$ $eV$, coherent results (blue dash line) corresponding to one-level model with energy $E_0-\mu=0.065$ $eV$, effect of the only electron-vibration coupling $E_P=0.018$ $eV$ (magenta dash-dot line), and effect of additional electron-electron interaction $U=0.3$ $eV$ (orange double dash-dot line).}
\label{Fig3}
\end{figure}

After the analysis of the effects of many-body interactions on the charge conductance and Seebeck coefficient, we can make a comparison with the experimental data available in \cite{Reddy2014} and shown in Figure (\ref{Fig1}). These data are plotted again in Figure (\ref{Fig3}) together with the fit discussed in Figure (\ref{Fig1}). We recall that for fullerene junctions the one-level model discussed in the previous section is characterized by the following parameters: $E_0-\mu=0.057$ $eV$, $\Gamma=0.032$ $eV$, and $\alpha=0.006$ $eV/V$. We remark that the level energy $\epsilon$ used in the previous discussion is related to the energy $E_0$ and the gate voltage $V_G$ through Equation (\ref{molevel}). Therefore, once fixed the value of $E_0$, one can switch from the energy 
$\epsilon$ to the gate voltage $V_G$. Before introducing many-body effects, we consider a slight shift of the level position considering the case  $E_0-\mu=0.065$ $eV$ reported in Figure (\ref{Fig3}). This energy shift is introduced to counteract the shifts of the peaks (conductance) or zeroes (Seebeck) introduced by many-body interactions which, in addition, reduce the amplitudes of response functions. The aim of this paper is to provide an optimal description for both charge conductance $G$ and Seebeck coefficient $S$. 

Starting from the level energy $E_0-\mu=0.065$ $eV$, in Figure (\ref{Fig3}), we analyze the effect of the electron-vibration coupling in the intermediate regime $E_P=0.018$ $eV$. The shift induced in the zero of the Seebeck coefficient is still compatible with experimental data. Moreover, the electron-vibration interaction shifts and reduces the peak of the charge conductance in an important way. However, this is still not sufficient to get an accurate description of the charge conductance.  One could increase the value of the coupling energy $E_P$, but, this way, the shift of the conductance peak becomes too large with a not marked reduction of the spectral weight.

Another ingredient is necessary to improve the description of both conductance $G$ and Seebeck coefficient $S$. In our model, the additional Coulomb repulsion plays a concomitant role. Its effects  poorly shift the zero of the Seebeck coefficient and slightly modifies the curve far from the zero. Therefore, the description of the Seebeck coefficient remains quite accurate as a function of the gate voltage. On the other hand, it provides a sensible reduction of the conductance amplitude with a not large shift of the peak. Hence, the effects of Hubbard term are able to improve the theoretical interpretation  of the experimental data for the conductance $G$ and the Seebeck coefficient $S$ close to the resonance. Far from the resonance, in a wider window of gate voltages, theory predicts the existence of a secondary peak of the conductance and a complex behavior of the Seebeck coefficient due to Coulomb blockade effects. The features are not negligible as a function of the gate voltage. 

\begin{figure}[t]
\centering
\includegraphics[width=10.0cm,angle=-90]{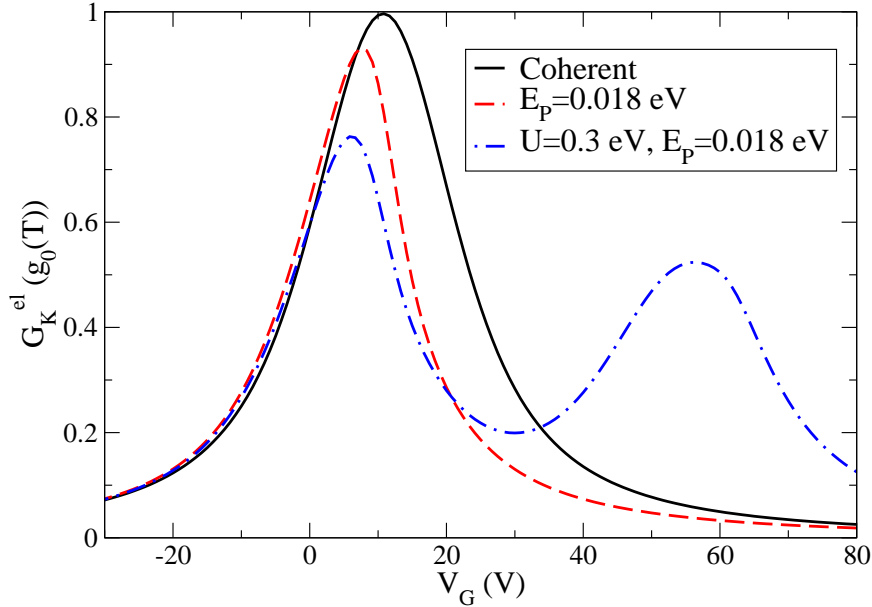}
\caption{Electronic thermal conductance $G_K^{el}$ in units of thermal conductance quantum $g_0(T)$  ($g_0(T) = \pi^2 k_{B}^{2} T/(3 h)$)  as a function the voltage gate $V_G$ in units of 
Volt at the temperature $T=100$ K:   coherent results (black solid line) corresponding to one-level model with energy $E_0-\mu=0.065$ $eV$, effect of the only electron-vibration coupling 
$E_P=0.018$ $eV$ (red dash line), and effect of additional electron-electron interaction $U=0.3$ $eV$ (blue dash-dot line). }
\label{Fig4}
\end{figure}

As far as we know, experimental measurements of the electronic thermal conductance $G_K^{el}$ have become only very recently available  \cite{Reddy2019}. Indeed, it is important to characterize this quantity since it allows to 
determine the thermoelectric figure of merit. Therefore, in Figure \ref{Fig4}, we provide the theoretical prediction of the electronic thermal conductance $G_K^{el}$ as a function of $V_G$ starting 
from the optimized values of the one-level parameters used to  describe both charge conductance and Seebeck coefficient in an accurate way. We stress that the plotted thermal conductance is expressed in terms of the  thermal conductance quantum $g_0(T) = \pi^2 k_{B}^{2} T/(3 h)$ \cite{Jezouin2013}. The main point is that, in the unities chosen in Figure \ref{Fig4}, the thermal conductance $G_K^{el}$ shows a strong resemblance with the behavior of the charge conductance $G$ in units of the conductance quantum $G_0$ as a function of the gate voltage $V_G$. We remark that, at $T=100$ K, $g_0(T) \simeq 9.456 \times 10^{-11} (W/K) \simeq 100 pW/K$. The values of the thermal conductance $G_K^{el}$ shown in Figure \ref{Fig4} are fractions of $g_0(T)$, therefore they are  fully compatible with  those estimated experimentally in hydrocarbon molecules \cite{Wang2007} (50pW/K).

\section{Conclusions} \label{Conclusions}
In this paper, we have theoretically analyzed the role of electron-vibration and electron-electron interactions on the thermoelectric properties of molecular junctions focusing on devices based on
fullerene.  We have used a self-consistent adiabatic approach which allows a non-perturbative treatment of the electron coupling to low frequency vibrational modes, such as those of the 
molecular center of mass between metallic electrodes. This approach incorporates Coulomb blockade effects due to strong electron-electron interaction between molecular degrees of freedom. 
We have analyzed a one-level model which takes into account the LUMO level of fullerene and its alignment to the chemical potential. We have stressed that an accurate description of the transport properties is obtained in the intermediate regime for the electron-vibration coupling  and in the strong coupling regime for the electron-electron interaction. Moreover, we have demonstrated that  only the combined effect of electron-vibration and electron-electron interactions is able to predict the correct behavior of both the charge conductance and the Seebeck coefficient. The theoretical calculations presented in this paper show a very good agreement with available experimental data of both charge conductance and Seebeck coefficient.

In this paper, we have used a one-level transport model as a starting point to address the role of many-body interactions between molecular degrees of freedom. This model is frequently used in all the cases where the energy levels 
can be tuned around the chemical potential and additional spectral features are absent \cite{Yee2011,Reddy2014}.  This is the case of the experiments in \cite{Reddy2014} for the fullerene junctions analyzed in this paper. The one-level model is expected to be valid for energies close to the Fermi level and low temperatures. Actually,  a more realistic description of the molecule and its coupling with metallic leads is needed if more complex transport phenomena take place, in particular interference effects \cite{Bai2019,Li2019} recently investigated in molecular junctions. Moreover, inclusion of quantum corrections to oscillator dynamics can be important in order to explore the effects of additional vibrational modes and further electron-vibration regimes \cite{Perroni_org} (from adiabatic to anti-adiabatic one) and their relation with strong electron-electron interactions \cite{Giulio,Perroni_Mott}.







\vspace{6pt} 

\end{document}